\newcommand{\roughly}[1]%
    {{\mathrel{\raise.3ex\hbox{$#1$\kern-.75em\lower1ex\hbox{$\sim$}}}}}
\newcommand{\lsim}{\mathrel{\roughly<}}
\newcommand{\gsim}{\mathrel{\roughly>}}

\documentclass[12pt]{article}
\usepackage{paper2e}
\usepackage{epsfig}
\usepackage{latexsym}

\begin{document}
\begin{titlepage}

\preprint{IPMU-09-0006}

\title{Ghost condensate and generalized second law}

\author{Shinji Mukohyama}

\address{
Institute for the Physics and Mathematics of the Universe (IPMU)\\ 
The University of Tokyo\\
5-1-5 Kashiwanoha, Kashiwa, Chiba 277-8582, Japan
}

\begin{abstract}
 Dubovsky and Sibiryakov recently proposed a scenario in which particles
 of different species propagate with different speeds due to their
 direct couplings to ghost condensate. It was argued that this extended
 version of ghost condensate allows a gedanken experiment leading to
 violation of the generalized second law. However, in the original ghost
 condensate scenario, difference in propagation speeds is suppressed
 by $M^2/M_{Pl}^2$, where $M$ is the order parameter of spontaneous
 Lorentz breaking and $M_{Pl}$ is the Planck scale. In this case the
 energy transfer necessary for the gedanken experiment is so slow that
 the timescale of decrease of entropy, if any, is always longer than the 
 Jeans timescale of ghost condensate. Hence the generalized second law
 is not violated by the gedanken experiment in the original ghost
 condensate scenario. This conclusion trivially extends to gauged ghost
 condensation by taking into account accretion of gauged ghost
 condensate into a black hole. 
\end{abstract}

\end{titlepage}

\section{Introduction}

Precision observational data recently revealed that the expansion of our
universe is accelerating. If Einstein's theory is correct, this requires
that more than $70$ \% of our universe is filled with invisible,
negative pressure, energy. This energy is named dark energy, but we do
not know what it really is. This situation reminds us of a story in the
nineteenth century: when the perihelion shift of Mercury was discovered,
some people hypothesized the existence of an invisible planet called
Vulcan, a so-to-speak dark planet, to explain the anomalous behavior of
Mercury. However, as we all know, the dark planet was not real and the
correct explanation was to change gravity, from Newton's theory to
Einstein's. With this in mind, it is probably natural to wonder if we
can change Einstein's theory at long distances to address the mystery of
dark energy.

From theoretical viewpoint, however, it is not easy to modify gravity in
infrared (IR). For example, massive gravity~\cite{Fierz:1939ix} and DGP
brane model~\cite{Dvali:2000hr} are known to have a ultraviolet (UV)
scale at around $1000$km, where effective field theories break
down~\cite{ArkaniHamed:2002sp,Luty:2003vm}. This implies that those
theories lose predictability at distances shorter than $1000$km. For the
DGP brane model in the branch without self-acceleration, nonlinear
effects provide an extra contribution to the kinetic term of the
longitudinal mode and push the UV scale to higher
energy~\cite{Nicolis:2004qq}. However, in the self-accelerating branch
this contribution has a wrong sign and, thus, this branch includes a
ghost.

Now let us remind ourselves of a situation with gauge field theory,
hoping to find a hint. In gauge field theory, simply adding mass to a
gauge boson changes the corresponding force law in IR but spoils its
well-behaved properties in UV. It is the Higgs mechanism that is useful
to modify the IR force law without ruining its UV behaviors. Indeed, the
Higgs mechanism is integral to the standard model of particle physics
and we are supposed to live in the Higgs phase of the theory.

Therefore, it seems promising to apply the idea of Higgs mechanism to
gravity to modify general relativity in IR. Ghost
condensation~\cite{ArkaniHamed:2003uy} is the simplest Higgs mechanism 
for gravity in the sense that it has only one Nambu-Goldstone
boson~\footnote{Closely related models include Lorentz-violating massive
gravity~\cite{Rubakov:2004eb,Dubovsky:2004sg}, Einstein aether 
theory~\cite{Jacobson:2000xp} and gauged ghost 
condensation~\cite{Cheng:2006us}. }. It opens up new type of
gravitational phenomenologies such as IR modification of 
gravity~\cite{ArkaniHamed:2003uy}, inflation with large 
non-Gaussianities~\cite{ArkaniHamed:2003uz,Senatore:2004rj,Cheung:2007st},
dark energy with
$w<-1$~\cite{Creminelli:2006xe,Mukohyama:2006be,Creminelli:2008wc}, rich
non-linear dynamics~\cite{ArkaniHamed:2005gu}, and so on.

The structure of the low energy effective field theory (EFT) of ghost
condensation is determined by the symmetry breaking pattern as in the
usual Higgs mechanism. We assume that (i) derivative of a scalar field
has a timelike constant vev, $\langle\partial_{\mu}\phi\rangle\ne 0$,
and that (ii) the background spacetime metric is maximally symmetric,
either Minkowski or de Sitter. By the assumption, the $4$-dimensional
spacetime diffeomorphism invariance is spontaneously broken down to the
$3$-dimensional spatial diffeomorphism invariance, i.e. the symmetry
under $\vec{x}\to\vec{x}'(t,\vec{x})$. Our strategy here is to write
down the most general action invariant under this residual
symmetry. After that, the action for the Nambu-Goldstone (NG) boson
$\pi$ is obtained by undoing the unitary gauge.  In Minkowski
background, the result is 
%
\begin{equation}
 L_{eff} =
  M^4\left\{\frac{1}{2}\left(\dot{\pi}-\frac{1}{2}h_{00}\right)^2 
      - \frac{\alpha}{M^2}(\vec{\nabla}^2\pi)^2 + \cdots
     \right\}, \label{eqn:effective-action}
\end{equation}
where $h_{00}$ is the time-time component of the metric perturbation,
$M$ is the scale of symmetry breaking and $\alpha$ ($>0$) is a
dimensionless constant of order unity. This low energy EFT is universal 
and should hold as far as the symmetry breaking pattern is the same. The
scale $M$ is the order parameter of spontaneous Lorentz breaking and
also plays the role of the UV cutoff scale of the low energy EFT. In the
$M/M_{Pl}\to 0$ limit, the $\pi$ sector is decoupled from gravity and 
general relativity is recovered. Also, proper analysis of 
scaling dimensions~\cite{ArkaniHamed:2003uy} shows that
higher-dimensional operators indicated by dots in
(\ref{eqn:effective-action}) are suppressed at least by some positive
(but in general fractional) power of $E/M$, where $E$ represents the
typical energy and/or momentum scale of the system.

If we include some of those higher-dimensional operators such as higher
time derivative terms in (\ref{eqn:effective-action}), then new modes
may appear but they are always outside the regime of validity of the
EFT: they have frequencies of the order $M$ or higher. Therefore, as far
as we are interested in physics at energies and momenta well below $M$,
those extra modes are irrelevant. Moreover, properties of those
high-frequency modes can be modified by other higher-dimensional
operators without any noticeable changes to low energy physics. For
these reasons we may and must concentrate on physics insensitive to
those higher-dimensional operators, i.e. physics at energies and momenta
well below $M$, unless we find a UV completion~\footnote{See
\cite{Graesser:2005ar,Mukohyama:2006mm} for some attempts towards a UV
completion.}. Note that the current phenomenological bound on $M$
is~\cite{ArkaniHamed:2005gu} 
%
\begin{equation}
 M\lsim 100GeV. \label{eqn:boundonM}
\end{equation}

The low energy effective action (\ref{eqn:effective-action}) plus the
Einstein-Hilbert action exhibits IR modification of Einstein's theory in
linearized gravitational potential~\cite{ArkaniHamed:2003uy}. The length 
scale $r_J$ and the timescale $t_J$ of the modification are given by 
%
\begin{equation}
 r_J \sim \frac{M_{Pl}}{M^2}
\end{equation}
and 
%
\begin{equation}
 t_J \sim \frac{M_{Pl}^2}{\sqrt{\alpha}M^3}. \label{eqn:Jeans-timescale}
\end{equation}
Note that both scales are much longer than $1/M$, provided that 
$M\ll M_{Pl}$. These scales are analogous to the Jeans scale and thus we
call them Jeans scales of ghost condensate.

Considering ghost condensation as a new candidate theory of gravity, it
is important to test its consistency. Black hole
thermodynamics~\cite{Mukohyama:1998ng} is probably useful for this
purpose.

The purpose of this paper is to investigate whether the generalized
second law of black hole thermodynamics holds in the presence of ghost
condensate. A doubt was raised recently by Dubovsky and
Sibiryakov~\cite{Dubovsky:2006vk}. They proposed an extended version of
ghost condensate in which particles of different species propagate with
different speeds due to their direct couplings to ghost condensate. It
was argued that Dubovsky-Sibiryakov's extension of ghost condensate
allows a gedanken experiment leading to violation of the generalized
second law. In this paper, on the contrary, we show that in the original
ghost condensate scenario the generalized second law cannot be violated
by the same gedanken experiment. The reason is that difference in
propagation speeds is suppressed by $M^2/M_{Pl}^2$, where $M$ is the
order parameter of spontaneous Lorentz breaking and $M_{Pl}$ is the
Planck scale, since Lorentz invariance recovers in the limit
$M^2/M_{Pl}^2\to 0$. We shall of course take into account direct
couplings generated by quantum corrections via gravitational
interactions.

The rest of this paper is organized as follows. In
Sec.~\ref{sec:DSextension} Dubovsky-Sibiryakov's extension of ghost
condensate is described and it is shown that a gedanken experiment
appears to violate the generalized second law in their theory. In
Sec.~\ref{sec:GSL} we go back to the original ghost
condensate scenario and show that the same gedanken experiment cannot
violate the generalized second law. In Sec.~\ref{sec:GGC} we consider
the gauged ghost condensate~\cite{Cheng:2006us} and show the same
result. Sec.~\ref{eqn:summary} is devoted to a summary of this paper and
some discussions.

\section{Dubovsky-Sibiryakov's extension}
\label{sec:DSextension}

Dubovsky and Sibiryakov~\cite{Dubovsky:2006vk} proposed an extension of
the ghost condensate scenario by adding a specific direct coupling
between the ghost condensate sector and matter fields. For a massless
scalar field $\psi$, they added a derivative coupling to the ghost
condensate sector as 
%
\begin{equation}
 S_{\psi} = \int \sqrt{-g}\left[
 -\frac{1}{2}g^{\mu\nu}\partial_{\mu}\psi\partial_{\nu}\psi
  + \frac{1}{2M_{DS}^4}
  \left(g^{\mu\nu}\partial_{\mu}\phi\partial_{\nu}\psi\right)^2
       \right]d^4x, \label{eqn:action-psi}
\end{equation}
where $\phi$ is the scalar field responsible for ghost condensate,
$M_{DS}$ is some energy scale. Typically, $M_{DS}$ should be of order
$M_{Pl}$ but we leave it as free in this extension.

The massless field $\psi$ described by the action (\ref{eqn:action-psi}) 
propagates with the speed different from that inferred from the light
cone structure of the metric $g_{\mu\nu}$. For example, in the Minkowski
background $g_{\mu\nu}=\eta_{\mu\nu}$ with $\phi=M^2t$, it is easy to
see that the propagation speed is not $1$ but $(1+\epsilon)^{-1/2}$,
where
%
\begin{equation}
 \epsilon = \frac{M^4}{M_{DS}^4}. 
\end{equation}
 Since $M_{DS}$ is typically of order $M_{Pl}$, $\epsilon$ is extremely
 small. Note that the known phenomenological upper bound on the order
 parameter of spontaneous Lorentz breaking in ghost condensate is
 (\ref{eqn:boundonM}). Nonetheless, we shall leave $\epsilon$ as a free
 parameter in this section.

For more general backgrounds with 
%
\begin{equation}
 g^{\mu\nu}\partial_{\mu}\phi\partial_{\nu}\phi = -M^4, 
  \label{eqn:background}
\end{equation}
the action (\ref{eqn:action-psi}) is rewritten as
%
\begin{equation}
 S_{\psi} = \int \sqrt{-g}\left[
 -\frac{1}{2}\tilde{g}^{\mu\nu}
 \partial_{\mu}\psi\partial_{\nu}\psi \right]d^4x
 = \int \sqrt{-\bar{g}}\left[
 -\frac{1}{2}\bar{g}^{\mu\nu}
 \partial_{\mu}\psi\partial_{\nu}\psi \right]d^4x,
 \label{eqn:action-psi-newmetric}
\end{equation}
where
%
\begin{eqnarray}
 \tilde{g}^{\mu\nu} & = & - (1+\epsilon) u^{\mu}u^{\nu}
  + (g^{\mu\nu}+u^{\mu}u^{\nu}), \nonumber\\
 \tilde{g}_{\mu\nu} & = & - \frac{1}{1+\epsilon}u_{\mu}u_{\nu}
  + (g_{\mu\nu}+u_{\mu}u_{\nu}), \nonumber\\
 \bar{g}_{\mu\nu} & = &
  \frac{\sqrt{-g}}{\sqrt{-\tilde{g}}}\times \tilde{g}_{\mu\nu},\nonumber\\
 \bar{g}^{\mu\nu} & = &
  \frac{\sqrt{-\tilde{g}}}{\sqrt{-g}}\times \tilde{g}^{\mu\nu},
\end{eqnarray}
and
%
\begin{equation}
 u_{\mu} = \frac{\partial_{\mu}\phi}{M^2}, \quad
  u^{\mu} = g^{\mu\nu}u_{\nu}.
\end{equation}
Note that $g^{\mu\nu}u_{\mu}u_{\nu}=-1$. The field $\psi$ propagates
along light cones of $\bar{g}_{\mu\nu}$ and its speed relative to
$u^{\mu}$ is not $1$ but again $(1+\epsilon)^{-1/2}$.

For a timescale sufficiently shorter than the Jeans timescale
(\ref{eqn:Jeans-timescale}), there is an approximate solution in 
Schwarzschild background~\cite{Mukohyama:2005rw}. The solution is as 
simple as 
%
\begin{equation}
 \phi = M^2\tau, \label{eqn:Sch-sol}
\end{equation}
where the Schwarzschild metric with the horizon radius $r_g$ is written
in the Lema\^{i}tre reference frame as
%
\begin{equation}
 ds^2 = -d\tau^2 + \frac{r_gdR^2}{r(\tau,R)} + r^2(\tau,R)d\Omega^2,
  \quad r(\tau,R) = \left[\frac{3}{2}\sqrt{r_g}(R-\tau)\right]^{2/3}. 
\end{equation}
This is not an exact solution but describes behavior of the system
at most up to the Jeans timescale (\ref{eqn:Jeans-timescale}). This
approximate solution indeed satisfies the condition
(\ref{eqn:background}). The corresponding effective metric
$\bar{g}_{\mu\nu}$ for $\psi$ turns out to be a Schwarzschild metric
with a different horizon radius $\bar{r}_g$, 
%
\begin{equation}
 d\bar{s}^2 = -d\bar{\tau}^2 +
  \frac{\bar{r}_gd\bar{R}^2}{\bar{r}(\bar{\tau},\bar{R})}
  + \bar{r}^2(\bar{\tau},\bar{R})d\Omega^2,
  \quad \bar{r}(\bar{\tau},\bar{R}) = 
  \left[\frac{3}{2}\sqrt{\bar{r}_g}(\bar{R}-\bar{\tau})\right]^{2/3},
\end{equation}
where 
%
\begin{equation}
 \bar{r}_g = (1+\epsilon)^{5/4}r_g, \quad 
  \bar{\tau} = (1+\epsilon)^{-1/4}\tau, \quad 
  \bar{R} = (1+\epsilon)^{-1/4}R.
\end{equation}
Therefore, while the original Schwarzschild metric $ds^2$ has 
temperature $T_{bh}=(4\pi r_g)^{-1}$, the $\psi$-metric $d\bar{s}^2$ has
a different temperature 
%
\begin{equation}
 T_{bh,\psi} = \frac{1}{4\pi\bar{r}_g}\times\frac{d\bar{\tau}}{d\tau}
  = \frac{T_{bh}}{(1+\epsilon)^{3/2}}.
\end{equation}
Here, temperatures $T_{bh}$ and $T_{bh,\psi}$ are defined with respect
to the original time variable $\tau$ at infinity and this is the reason
why the factor $d\bar{\tau}/d\tau$ is included in the above expression
for $T_{bh,\psi}$. Since different fields can have different $\epsilon$,
a black hole can have different temperatures for different species.

The gedanken experiment in the Dubovsky-Sibiryakov extension consists of
a Schwarzschild black hole with horizon radius $r_g$, two species $A$
and $B$ with $\epsilon_A>\epsilon_B$, and two spherical static shells
surrounding the black hole each of which is made of $A$ and $B$,
respectively. We denote temperatures of the black hole for $A$ and $B$
as $T_{bh,A}$ and $T_{bh,B}$ ($T_{bh,A}<T_{bh,B}$) and suppose that the
shells of $A$ and $B$ have temperatures $T_{shell,A}$ and $T_{shell,B}$,
respectively. One can tune the shells' temperatures so that 
%
\begin{equation}
 T_{bh,A} < T_{shell,A} < T_{shell,B} < T_{bh,B}
  \label{eqn:temperatures}
\end{equation}
and that
%
\begin{equation}
 F_{shell\to bh,A} = - F_{shell\to bh,B} > 0,
  \label{eqn:energy-balance}
\end{equation}
where 
%
\begin{equation}
 F_{shell\to bh,i} = \frac{\pi^3r_g^2}{15}
  \left[\Gamma_i(T_{shell,i})T_{shell,i}^4-\Gamma_i(T_{bh,i})T_{bh,i}^4\right]
  \quad (i=A,B)
  \label{eqn:netflux}
\end{equation}
is the net flux of energy from the shell to the black hole for the species
$i$ and $\Gamma_i(T)=O(1)$ is a slowly varying function representing the
gray body factor for the species $i$.

Note that the action (\ref{eqn:action-psi-newmetric}) is exactly the
same as the usual canonical action for a massless scalar field
propagating in the corrected metric $d\bar{s}^2$ (not the original
metric $ds^2$). Therefore, following the standard quantization procedure
for a canonical scalar field in a fixed background geometry, we obtain
the usual formula (\ref{eqn:netflux}) for the net energy flux. Of
course, treating the geometry $d\bar{s}^2$ as a fixed background is just
an approximation. This approximation is justified if (i) the
backreaction of Hawking radiation to the geometry is small and if (ii)
relevant processes in the gedanken experiment are sufficiently faster than
non-trivial dynamics of ghost condensate (such as Jeans-like instability
and/or accretion into black hole). The condition (i) is satisfied if the
black hole is sufficiently large. We shall investigate the condition
(ii) in the next section.

To be more precise, (\ref{eqn:netflux}) should be understood as the net
flux of gravitational energy for each species $i$. Note that in the
presence of direct couplings to ghost condensate, gravitational energy
and particle-physics energy are not the same in general. For $\psi$ in
flat background ($g_{\mu\nu}=\eta_{\mu\nu}$ with $\phi=M^2t$), 
gravitational energy density, or $T_{00}$, is 
%
\begin{equation}
 \rho_{grav} = T_{00} = \frac{1}{2}(1+3\epsilon)\dot{\psi}^2 +
  \frac{1}{2}(\vec{\nabla}\psi)^2, \nonumber\\
\end{equation}
while the particle-physics energy density, or the Hamiltonian density, is 
%
\begin{equation}
 \rho_{part} = {\cal H} = \frac{1}{2}(1+\epsilon)\dot{\psi}^2 + 
  \frac{1}{2}(\vec{\nabla}\psi)^2. 
\end{equation}
For small $\epsilon$, the difference is $O(\epsilon)$, as in
(\ref{eqn:ratio-energies}) below.

In this setup, for a timescale sufficiently shorter than the Jeans
timescale $t_J$, the black hole mass does not change. Energy is just
transferred via the black hole from the shell of $A$ with lower
temperature $T_{shell,A}$ to the shell of $B$ with higher temperature
$T_{shell,B}$. Thus, the sum of entropies of two shells decreases. On
the other hand, if we consider the horizon area (in the Planck unit)
divided by four as black hole entropy then the black hole entropy does
not change. In this way, the total entropy appears to decrease. Thus,
this gedanken experiment in the Dubovsky-Sibiryakov extension appears to
violate the generalized second law.

\section{No violation of GSL in the original ghost condensate}
\label{sec:GSL}

It is widely believed that the generalized second law should hold at
least for quasi-stationary evolution of systems with black holes. In
situations which have holographic descriptions, a black hole is dual to
a thermal excitation and, thus, the generalized second law is dual to 
the ordinary second law of thermodynamics. Therefore, violation of the
generalized second law would indicate lack of holographic
descriptions~\cite{ArkaniHamed:2007ky}. For this reason, by demanding
validity of the generalized second law, we may hope to exclude regions
in the parameter space that do not allow holographic dual descriptions.

In the previous section we have reviewed the gedanken experiment in the
Dubovsky-Sibiryakov extension of ghost condensate. However, we have not
specified the scale $M_{DS}$ (and thus the value of $\epsilon$). Also,
we have not taken into account two important time scales associated with
the gedanken experiment. One is the Jeans time scale
(\ref{eqn:Jeans-timescale}) of ghost condensate, and the other is the
time scale in which shells' entropy would decrease. If the latter time
scale is longer than the former then the gedanken experiment is
invalidated. In this section we shall show that this is indeed the case 
in the original proposal of ghost
condensation~\cite{ArkaniHamed:2003uy,ArkaniHamed:2005gu}. 
In this case, excitations of ghost condensate become important before
the gedanken experiment starts operating. Those excitations should
accrete into the black hole and, accordingly, the total entropy is
expected to increase.

In the original ghost condensate the scale $M$ is the order parameter of
spontaneous Lorentz breaking. Therefore, Lorentz invariance should
recover in the limit $M^2/M_{Pl}^2\to 0$ limit. In particular,
$\epsilon$ should vanish in this limit and thus we have 
%
\begin{equation}
  \epsilon = O\left(\frac{M^2}{M_{Pl}^2}\right).
   \label{eqn:epsilonsmall}
\end{equation}
Note that the typical value of $M_{DS}$ is of order $M_{Pl}$ and leads
to $\epsilon=O(M^4/M_{Pl}^4)$. This corresponds to the typical strength
of direct couplings generated by quantum corrections via gravitational
interactions. The condition (\ref{eqn:epsilonsmall}) is less restrictive
than this, but is still more than sufficient to protect the generalized
second law.

By setting 
%
\begin{equation}
  \epsilon_i = O\left(\frac{M^2}{M_{Pl}^2}\right),
\end{equation}
we obtain
%
\begin{equation}
 T_{bh,i} = T_{bh}\times \left[ 1 + O(\epsilon_i) \right],
\end{equation}
and
%
\begin{equation}
 T_{bh,B}-T_{bh,A} = T_{bh}\times O\left(\frac{M^2}{M_{Pl}^2}\right). 
\end{equation}
Since $T_{shell,i}$ are bounded from below and from above by
$T_{bh,A}$ and $T_{bh,B}$, respectively as in (\ref{eqn:temperatures}),
it also follows that 
%
\begin{equation}
 T_{shell,i}-T_{bh,i} = T_{bh}\times O\left(\frac{M^2}{M_{Pl}^2}\right). 
\end{equation}
This implies that the net flux of energy for each species is 
%
\begin{equation}
 |F_{shell\to bh,i}| \sim r_h^2T_{bh}^3\times |T_{shell,i}-T_{bh,i}|
  = T_{bh}^2\times O\left(\frac{M^2}{M_{Pl}^2}\right). 
\end{equation}

Since the heat transfer is suppressed by $M^2/M_{Pl}^2$, the rate of
decrease of shells' entropy is also suppressed. Of course, if we could
wait for infinite time then the gedanken experiment would still lead to 
violation of the generalized second law. However, as already stated, the
setup of the gedanken experiment does not persist forever but can last
only up to the Jeans timescale (\ref{eqn:Jeans-timescale}) at
most. Thus, under the equi-flux condition (\ref{eqn:energy-balance}),
let us estimate the maximum decrease of shells' entropy which the
gedanken experiment of this sort could in principle lead to within the
Jeans timescale $t_J$:
%
\begin{eqnarray}
 |\Delta S_{shells}|_{max} & = & 
  \left| \frac{dS_{shell,A}}{dt}+\frac{dS_{shell,B}}{dt}
  \right| \times t_J \nonumber\\
 & = & 
  F_{shell\to bh,A}\times 
  \left(\frac{1}{T_{shell,A}}
   \cdot\frac{dE_{part,A}}{dE_{grav,A}}
  - \frac{1}{T_{shell,B}}
  \cdot\frac{dE_{part,B}}{dE_{grav,B}}\right)
  \times t_J \nonumber\\
 & \sim &
  O\left(\frac{M^2}{M_{Pl}^2}\right) \times
  \frac{F_{shell\to bh,A}}{T_{bh}}\times t_J \nonumber\\
 & \sim &
  O\left(\frac{M^2}{M_{Pl}^2}\right) \times \frac{T_{bh}}{M}, 
  \label{eqn:DeltaSmax}
\end{eqnarray}
where 
%
\begin{equation}
 \frac{dE_{part,i}}{dE_{grav,i}} = 1 + O(\epsilon_i)
  \label{eqn:ratio-energies}
\end{equation}
is the ratio of particle-physics energy to gravitational energy for the
species $i$.

Validity of the effective field theory requires that the black hole
temperature be lower than the UV cutoff scale $M$: 
%
\begin{equation}
 T_{bh} \ll M. \label{eqn:Tbhmax}
\end{equation}
Equivalently, the horizon radius $r_g$ must be sufficiently greater than
$1/M$. Otherwise, we cannot trust the low energy effective field 
theory and cannot justify the gedanken experiment at all. Under this
condition, the maximum decrease of shells' entropy (\ref{eqn:DeltaSmax})
is bounded as 
%
\begin{equation}
 |\Delta S_{shells}|_{max} \ll O\left(\frac{M^2}{M_{Pl}^2}\right) 
  \ll O(1). 
\end{equation}
Therefore, shells' entropy cannot decrease even by $O(M^2/M_{Pl}^2)$ 
within the Jeans timescale of ghost condensate. In other words, the
timescale of decrease of shells' entropy, if any, is always longer than
the maximum timescale for which the setup of the gedanken experiment can
in principle last.

In conclusion, the gedanken experiment suggested in
\cite{Dubovsky:2006vk} does not violate the generalized second law in
the original ghost condensation scenario. The essential reason for this
is that difference in propagation speeds is suppressed by
$M^2/M_{Pl}^2$, where $M$ is the order parameter of spontaneous Lorentz 
breaking and $M_{Pl}$ is the Planck scale, since Lorentz invariance
recovers in the limit $M^2/M_{Pl}^2\to 0$. This makes the semiclassical
heat flow so slow that shells' entropy cannot decrease before accretion
of ghost condensate induced by the Jeans instability increases black
hole entropy.

\section{Gauged ghost condensate}
\label{sec:GGC}

As another example, let us consider gauged ghost
condensation~\cite{Cheng:2006us}. In this case, if the gauge coupling is
large enough then the Jeans instability disappears. Nonetheless, there
still is the maximum time scale in which the gedanken experiment can in 
principle last. In the presence of the higher derivative term, the black
hole solution (\ref{eqn:Sch-sol}) is just an approximate
solution. Actually, gauged ghost condensate slowly gets excited and
those excitations accrete towards the black hole. As a result, the black
hole entropy increases. The time scale of accretion $t_{acc}$ can be
estimated from eq.~(6.13) of \cite{Cheng:2006us}. The result is
%
\begin{equation}
 t_{acc} \sim \frac{m_{bh}M_{Pl}}{M^3} 
  \sim t_J \times \frac{M_{Pl}}{T_{bh}},  
\end{equation}
where we have defined $t_J$ by (\ref{eqn:Jeans-timescale}). Since
$T_{bh}\ll M\ll M_{Pl}$ is required for validity of the effective field
theory, we have $t_{acc}\gg t_J$. Thus, the absence of Jeans instability
makes it possible for the gedanken experiment in the gauged ghost
condensate to last longer than in the (ungauged) ghost condensate.

Nonetheless, assuming again that $\epsilon_i=O(M^2/M_{Pl}^2)$ and
repeating the analysis in the previous section, we still obtain 
%
\begin{equation}
 \left| \Delta S_{shells}\right|_{max}
  \sim O\left(\frac{M}{M_{Pl}}\right)
  \ll O(1). 
\end{equation}
In other words, black hole entropy increases due to accretion before 
Dubovsky-Sibiryakov's gedanken experiment starts operating. Thus, the
gedanken experiment does not lead to violation of the generalized second
law in gauged ghost condensation.

\section{Summary and discussion}
\label{eqn:summary}

We have revisited the gedanken experiment suggested by Dubovsky and
Sibiryakov~\cite{Dubovsky:2006vk}. In their extension of ghost
condensate, the gedanken experiment appears to violate the generalized
second law. This result may imply lack of holographic dual descriptions 
for their extension~\cite{ArkaniHamed:2007ky}.

Dubovsky-Sibiryakov's extension requires difference in propagation
speeds for different species. In the limit where Lorentz invariance is
recovered, the generalized second law should recover.

In the original ghost condensate scenario, difference in propagation
speeds is suppressed by $M^2/M_{Pl}^2$, where $M$ is the order 
parameter of spontaneous Lorentz breaking due to ghost condensation and
$M_{Pl}$ is the Planck scale. This is because Lorentz invariance
recovers in the limit $M^2/M_{Pl}^2\to 0$. For this reason, the energy 
transfer necessary for the gedanken experiment is so slow that the
timescale of decrease of shells' entropy, if any, is always longer than
the Jeans timescale of ghost condensate. The latter is the maximum
timescale for which the setup of the gedanken experiment can in
principle last. Therefore, the gedanken experiment does not lead to
violation of the generalized second law in the original ghost
condensation scenario. Note that the Jeans timescale is longer than the
present age of the universe if $M\lsim 10MeV$. Even in this case the
Jeans time is not long enough for the gedanken experiment.

We have also shown a similar result for the gauged ghost
condensation. In this case, we have considered the time scale of black
hole accretion and shown that the gedanken experiment cannot decrease
the total entropy within this time scale. It is also possible to apply
this consideration to the ungauged ghost condensation since the
accretion rate is the same. However, the argument in Sec.~\ref{sec:GSL} 
based on Jeans timescale suffices in the ungauged case.

Jeans instability of ghost condensate disappears in de Sitter background 
if $H\gsim 1/t_J$, where $H$ is the Hubble expansion rate and $t_J$ is
the Jeans timescale. Hence, one might expect that the setup of the
gedanken experiment could last longer than $t_J$ in asymptotically de
Sitter backgrounds, say Schwarzschild-de Sitter background. However,
this is most likely too naive. The essential reason for the absence of
Jeans instability in a pure de Sitter background (with large enough $H$)
is that the vev $\langle\partial_{\mu}\phi\rangle$ has a positive
expansion. On the other hand, in a Schwarzschild-de Sitter background,
the vev $\langle\partial_{\mu}\phi\rangle$ has a negative expansion
(i.e. it is contracting) near the black hole, while it has a positive
expansion far enough from the black hole. Therefore, while the Jeans
instability can disappear in the far region, it should exist near the
black hole. In particular, in the intermediate region where the
expansion almost vanishes, the Jeans instability should be essentially
the same as that in flat background. For this reason, in either
asymptotically flat or asymptotically de Sitter background, the Jeans 
timescale estimated in Minkowski background is the maximum timescale for
which the setup of the gedanken experiment can in principle
last. Therefore, even in asymptotically de Sitter background, the
gedanken experiment does not violate the generalized second law in the
original ghost condensate scenario.

Now let us discuss extensions of the gedanken experiment to get some
insights about UV completion of ghost condensate.

So far, we have considered only two species $A$ and $B$ in the gedanken
experiment. What happens if we consider many species with different
$\epsilon_i$ and thus with different propagation speeds? Of course, only
light degrees of freedom can contribute to the gedanken experiment: they
must be sufficiently lighter than $T_{bh}$. Based on the gedanken
experiment with many light degrees of freedom, we conjecture that 
%
\begin{equation}
 \frac{1}{\sqrt{\alpha}}\sum_{1\leq i<j\leq N} (\epsilon_i-\epsilon_j)^2
  = O\left(\frac{M^2}{M_{Pl}^2}\right)
\end{equation}
should be satisfied by a sensible UV completion. Here, $M$ is the order
parameter of ghost condensation, $\alpha$ ($>0$) is the dimensionless
coefficient of the relevant higher derivative term (see
(\ref{eqn:effective-action})) and $N$ is the number of light fields
whose mass is well below $M$. If this conjecture is correct then the
extended gedanken experiment does not violate the generalized second
law.

Next let us consider extension involving ghost condensate quanta. So
far, we have treated ghost condensate as a fixed background and have not
considered its excitation $\pi$ ($\propto\delta\phi$). A natural
question now is ``what happens if we replace the species $A$ or $B$ in
Dubovsky-Sibiryakov's gedanken experiment by $\pi$?'' To address this
question let us first consider how a black hole radiates $\pi$ quanta in
the absence of higher-derivative terms. Since the rest frame of the
ghost condensate background is in-falling towards the black hole and
$\pi$ has a vanishing sound speed, $\pi$ quanta cannot escape to
infinity. This means that there would be no radiation of $\pi$ quanta
from a black hole in the absence of higher-derivative terms. In reality,
higher-derivative terms are present and a black hole radiates $\pi$
quanta. The spectrum of Hawking radiation of $\pi$ is still highly
suppressed and non-thermal~\cite{Feldstein:2008yx} although the result
seems UV sensitive. From this, one might naively guess that by replacing
the species $A$ (the one with lower temperature) in the gedanken
experiment by $\pi$, the generalized second law could be
violated. However, this is not necessarily true. In order to start this
modified gedanken experiment, we need to prepare a quasi-static shell
made of thermally excited $\pi$ quanta. The precise way the shell is
prepared may be UV sensitive (because of e.g. formation of
caustics~\cite{ArkaniHamed:2005gu}) but should be through gravity
anyway. Let us assume that a quasi-static shell is somehow prepared and
ask how it radiates. In the absence of higher-derivative terms, it would
not radiate $\pi$ quanta since $\pi$ has vanishing sound speed. In the
presence of higher-derivative terms, the thermal shell of $\pi$ can
radiate but the spectrum of $\pi$ radiation is most likely non-thermal
and highly suppressed. We now conjecture that {\it a sensible UV
completion should be such that the spectrum of $\pi$ radiation from a
black hole and the spectrum from a thermal shell of $\pi$ with the same
temperature are essentially identical}~\footnote{Difference in the two
spectra is allowed to the extent that the timescale of decrease of total
entropy, if any, is longer than either the Jeans timescale of ghost
condensate or the lifetime of the quasi-static shell.}. If this
conjecture is correct then energy flows stop when all three temperatures
agree and the total entropy is maximized. Thus the conjecture is 
sufficient to prevent the generalized second law from being violated by
this modified gedanken experiment. Note that this conjecture does not
contradict with any known facts. While the proof (or disproof) of this
conjecture requires a concrete setup for UV completion and is beyond the
scope of the present paper, it is appropriate to state it as a
conjecture since there is no known contradiction.

In this paper we have not considered yet-another gedanken experiment
suggested by Eling, et al.~\cite{Eling:2007qd}. It is based on a purely 
classical process analogous to the Penrose
process~\cite{Penrose:1969pc,Penrose:1971uk}. They mainly considered
Einstein aether theory~\cite{Jacobson:2000xp} but the process could in
principle be applied to Dubovsky-Sibiryakov's extension of ghost
condensate (with relatively large $|\epsilon_i|$). They concluded that
the generalized second law can be violated in these theories. However,
in the original ghost condensate scenario, $|\epsilon_i|$ is generated
by quantum corrections and thus suppressed by $M^2/M_{Pl}^2$. In this
case one can show that the analogue of the Penrose process is
kinematically forbidden unless particles are initially released from
positions extremely close to the horizon. The essential reason for this
is that the ``ergo-region'' disappears in the $\epsilon_i\to 0$
limit. This kinematical fact means that particles just before being
released are accelerated rather strongly and quantum effects such as
buoyancy force due to the thermal bath near the horizon may become
important. Actually, as Unruh and Wald showed in
\cite{Unruh:1982ic,Unruh:1983ir} (see also
\cite{Shimomura:1999xp,Gao:2001ut}.), the buoyancy force is essential
for recovery of the generalized second law in a gedanken experiment in 
general relativity. At the very least, it is probably fair to say that
we should include quantum effects consistently~\footnote{If we do not
take into account quantum effects at all then $\epsilon_i=0$ in the
original ghost condensate scenario and there is no paradox in the first
place.}. Also, we may have to take into account the fact that not only
the analogue of the Penrose process but also normal processes are
allowed in the ``ergo-region''. These normal processes of course include
e.g. soft elastic scattering and can dominate over the analogue of the
Penrose process, especially in the thin ``ergo-region'' limit. Careful
consideration of these issues is an interesting future subject but
beyond the scope of this paper.

Excitations of ghost condensate can carry not only positive energy but
also negative energy. By sending negative energy to a black hole, one
might hope to violate the generalized second
law~\cite{Feldstein:2009qy}. However, the process considered in 
\cite{Feldstein:2009qy} does not satisfy basic conservation
laws. Gravitational energy of excitation of ghost condensate is the sum
of particle-physics energy and the charge associated with shift
symmetry~\cite{ArkaniHamed:2003uy}. The particle-physics energy is
non-negative and the shift charge is conserved. Hence, excitations with
negative gravitational energy are always accompanied by excitations with
larger positive energy. Therefore, if a negative energy falls into a
black hole then more positive energy follows after that. This means that
the black hole entropy should increase after all, if the conservation law
is properly taken into account.

\section*{Note added}

In Sec.~\ref{eqn:summary} we have briefly commented on a yet-another
gedanken experiment suggested by Eling, et al.~\cite{Eling:2007qd}. In
ref.~\cite{Mukohyama:2009um} it was shown that efficiency of the
gedanken experiment is highly suppressed by the factor $M^2/M_{Pl}^2$
and that it is always lower than accretion of ghost condensate into a
black hole. For this reason, the gedanken experiment suggested by Eling,
et al.~\cite{Eling:2007qd} does not violate the generalized second law
in ghost condensate. In Sec.~\ref{eqn:summary} we have also commented on 
negative energy carried by excitations of ghost condensate. In this
respect, ref.~\cite{Mukohyama:2009um} proved an averaged null energy
condition, which prevents the negative energy from violating the
generalized second law in a coarse-grained sense.

\section*{Acknowledgements}

The author would like to thank Hsin-Chia Cheng, Sergei Dubovsky, Andrei
Linde, Sergey Sibiryakov, Takahiro Tanaka and Aron Wall for useful
comments. The work of the author was supported in part by MEXT through a
Grant-in-Aid for Young Scientists (B) No.~17740134, by JSPS through a
Grant-in-Aid for Creative Scientific Research No.~19GS0219, and by the
Mitsubishi Foundation. This work was supported by World Premier
International Research Center Initiative¡ÊWPI Initiative), MEXT, Japan.


\end{document}